\documentclass[twocolumn,showpacs,amsmath,amssymb,aps,prl,superscriptaddress]{revtex4}
\usepackage{graphicx}     
\usepackage{bm}
\begin{document}

\title{Scaling Behavior of Entanglement in Two- and Three-Dimensional
Free Fermions}
\author{Weifei Li}
\affiliation{Department of Physics and Astronomy, University of
Southern California, Los Angeles, CA 90089}
\author{Letian Ding}
\affiliation{Department of Physics and Astronomy, University of
Southern California, Los Angeles, CA 90089}
\author{Rong Yu}
\affiliation{Department of Physics and Astronomy, University of
Southern California, Los Angeles, CA 90089}
\author{Tommaso Roscilde}
\affiliation{Department of Physics and Astronomy, University of
Southern California, Los Angeles, CA 90089}
\affiliation{Max-Planck-Institut f\"ur Quantenoptik,
Hans-Kopfermann-strasse 1, 85748 Garching, Germany}
\author{Stephan Haas}
\affiliation{Department of Physics and Astronomy, University of
Southern California, Los Angeles, CA 90089}

\date{\today}

\begin{abstract}

Exactly solving a
spinless fermionic system in two and three dimensions,
we investigate the scaling behavior of
the block entropy in critical and non-critical phases.
The scaling of the block entropy crucially depends on the
nature of the excitation spectrum of the system and
on the topology of the Fermi surface. Noticeably,
in the critical phases the scaling violates the area
law and acquires a logarithmic correction \emph{only} when
a well defined Fermi surface exists in the system.
When the area law is violated, we accurately verify a conjecture
for the prefactor of the logarithmic correction, proposed by
D. Gioev and I. Klich [quant-ph/0504151].
\end{abstract}

\maketitle

The nature of many-body
entanglement in various solid-state models has been the focus
of recent interest. The motivation for this
effort is two-fold. On the one side, these systems are of
interest for the purpose of quantum information processing and quantum
computation \cite{Burkard06}.
At a fundamental level, the study of entanglement represents
a \emph{purely quantum} way of understanding and characterizing
quantum phases and quantum phase transitions
in many-body physics \cite{Osterlohetal02, OsborneN02,
Verstraeteetal04,Vidaletal03}.

A striking feature of entangled states $|\Psi\rangle$ is that
a \emph{local} accurate description of such states is impossible,
namely each subsystem $A$ of the total system $U$ can have a
\emph{finite} entropy, quantified as the von-Neumann entropy
$S_A = -{\rm Tr} \rho_A \log_2 \rho_A$  of its reduced
density matrix $\rho_A = Tr_{U\setminus A} |\Psi\rangle \langle \Psi |$,
whereas the total system clearly has zero
entropy. The \emph{entropy of entanglement} $S_A$ of the subsystem is
a reliable estimate of the entanglement between the subsystem $A$
and the rest, $U\setminus A$. Assuming that the system $U$
corresponds to the whole universe in $d$ dimensions, a fundamental question
concerns the scaling behavior of the entropy of entanglement $S_L$
of an hypercubic subsystem $L^{d}$ (hereafter denoted as a \emph{block})
with its size $L$. Indeed,
such scaling probes directly the spatial range of entanglement:
when the block size exceeds the characteristic length
over which two sites are entangled, the block
entropy should become subadditive, and scale at
most as the area of the block boundaries, following
a so-called \emph{area law}: $S_L \sim L^{d-1}$.
A crucial question is then if and how the scaling of
the block entropy changes when the nature of
the quantum many-body state evolves in a critical way
by passing through a quantum phase transition,
and how the characteristic spatial extent of entanglement
relates to the correlation length of the system.

 This question has been extensively addressed
in the case of one-dimensional spin systems
\cite{Vidaletal03,Latorreetal04,RefaelM04,Laflorencie05}, in chains
of harmonic oscillators \cite{Skrovseth05, Crameretal06} and
in related conformal field theories (CFT) \cite{CalabreseC04,Korepin04}.
Here it is found unambiguosly that
in states with exponentially decaying (connected) correlators
$S_L$ follows the area law $S_L\sim L^0$, saturating to a finite value,
whereas for critical states, displaying
power-law decaying correlations, a \emph{logarithmic correction}
to the area is always present: $S_L = [(c+\bar{c})/6] \log_2 L$,
where $c$ is the central charge of the related CFT.
The asymptotic value of the block entropy is
found to diverge logarithmically with the correlation length,
$S_{\infty} \sim \log_2(\xi)$, which clearly establishes
the relationship between entanglement and correlations.
The above picture holds true only in presence of
\emph{short-range} interactions; on the contrary, in presence
of long-range interactions the divergence of the
correlation length can be still accompanied by the
area law \cite{Dueretal05,Crameretal06}.

 In higher spatial dimensions less results are available,
and the general relationship between the block entropy scaling
and the correlation properties of the quantum state
is still unclear even for short-range interactions.
In free-boson systems, it has been generally proven
that the area law is satisfied for non-critical systems
\cite{Plenioetal05,Crameretal06}.
For free-fermion systems, on the other hand, it has
been proven \cite{Wolf06,GioevK05} that critical systems
with short-range hoppings
and a finite Fermi surface exhibit a logarithmic correction
to the area law
\begin{equation}
S_L = C/3 ~(\log_2 L)~ L^{d-1}.
\label{e.logcorr}
\end{equation}

In this paper, we investigate a general quadratic fermionic Hamiltonian
both in $d=2$ and $d=3$. Upon tuning the Hamiltonian parameters,
this model has distinct critical phases \emph{with} and
\emph{without} a finite Fermi surface, as well as non-critical
phases. The scaling behavior of the block entropy is accurately
obtained through exact diagonalization. In non-critical
states we find that the area law indeed holds, and
we confirm that logarithmic corrections to such law
are present in critical states with a finite Fermi
surface, as found in Refs. \onlinecite{Wolf06,GioevK05}.
The prefactor $C$ of the $L$-dependence of $S_L$
in Eq.(\ref{e.logcorr}) is found to be very accurately
predicted by a formula based on the Widom conjecture
\cite{GioevK05}. On the other hand, for critical states with a Fermi
surface of zero measure, we find that the corrections
to the area law are either \emph{absent} or \emph{sublogarithmic}.
This means that the relationship between entanglement
and correlations in higher dimensional systems is
different than in $d=1$, and that a crucial role
is played by the geometry of the Fermi surface
or, alternatively, by the density of states at the ground
state energy.


We consider a bilinear spinless fermionic system
on a $d$-dimensional hypercubic lattice with
hopping and pairing between nearest-neighbor lattice sites
\begin{eqnarray}
H=\sum_{\langle{\bm i}{\bm j}\rangle} \left[ c^{\dagger}_{{\bm i}}c_{{\bm j}}-\gamma
(c^{\dagger}_{{\bm i}}c^{\dagger}_{{\bm j}}+c_{{\bm j}}c_{{\bm i}})\right] -
\sum_{{\bm i}}2\lambda c^{\dagger}_{{\bm i}}c_{{\bm i}}. \label{e.Hamilton}
\end{eqnarray}
 $\lambda$ is the chemical
potential, while $ \gamma$ is the pairing potential. The sum of
$\sum_{\langle {\bm i} {\bm j}\rangle} $ extends over nearest-neighbor pairs.
The above Hamiltonian is a $d>1$ generalization of the 1$d$
spinless fermionic Hamiltonian which is obtained by
Jordan-Wigner transformation of the XY model in a transverse field
\cite{LiebSM61}. Although in $d>1$ the exact relationship
to the XY model is lost, we can imagine the above Hamiltonian
to represent the effective fermionic degrees of freedom
of an interacting system with quantum-critical phases.

A more insightful expression for the Hamiltonian of
Eq. (\ref{e.Hamilton}) is obtained upon
Fourier transformation to momentum space:
\begin{eqnarray}
H&=&\sum_{\bm k} \left[ - 2t_{\bm k} c^{\dagger}_{\bm k}c_{\bm k}
 + i\Delta_{\bm k}
 (c^{\dagger}_{\bm k}c^{\dagger}_{-{\bm k}}+c_{-{\bm k}}c_{\bm k})\right]
 \nonumber \\
&& t_{\bm k} = \lambda - \sum_{\alpha=1}^{d}\cos k_{\alpha} ~~~~~
 \Delta_{\bm k} = \gamma \sum_{\alpha=1}^{d}\sin k_{\alpha}
\end{eqnarray}
The pairing potential in $\bm k$-space, $\Delta_{\bm k}$,
clearly reveals a $p$-wave symmetry.

\begin{figure}
\includegraphics[width=3.2in]{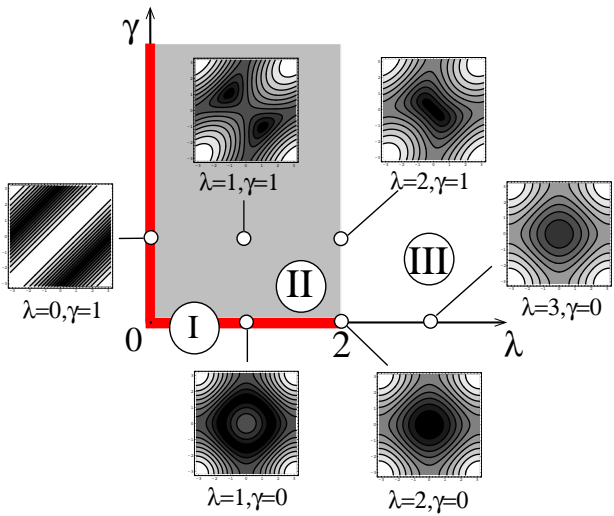}
\caption{Phase diagram of the model Eq.(\ref{e.Hamilton}) for the
case $d=2$. The roman numbers for the various phases are explained
in the text. Representative contour plots of the dispersion relation
$\Lambda_{\bm k}$ are also shown. There the black areas corresponds
to $\Lambda_{\bm k}=0$ and the white areas to the top of the band.}
 \label{f.phd}
\end{figure}

 This Hamiltonian can be diagonalized exactly by
a Bogoliubov transformation, to give
\begin{equation}
H = \sum_{\bm k}\Lambda_{\bm k}f^{+}_{\bm k}f_{\bm k}
~~~~~~ \Lambda_{\bm k} = 2\sqrt{ t_{\bm k}^2 + \Delta_{\bm k}^2}
\end{equation}
Depending on the parameters $\gamma$ and $\lambda$, this system has
a rich phase diagram, including metallic, insulating and ($p$-wave)
superconducting regimes, as shown in Fig. \ref{f.phd}.
The different phases are certainly distinguished
by the different decay of the correlation function,
which tells apart the critical from the non-critical phases.
Nonetheless, a classification which turns out to be
relevant for the study of entanglement is based on
the features of the gapless
excitation manifold $\Lambda_{\bm k}=0$. Such manifold
can be characterized by the
\emph{density of states} at the ground-state energy $g(0)$,
and by the
so-called \emph{co-dimension} \cite{Volovik03,Volovik05}
$\bar{d}$, defined as the dimension
of ${\bm k}$-space minus the dimension of the $\Lambda_{\bm k} = 0$
manifold.
We notice that the existence of a finite Fermi surface at zero energy
implies that $\bar{d}=1$ and $g(0)>0$, while in absence of
a finite Fermi surface we have $\bar{d}\geq 2$ and
$g(0)>0$ or $g(0)=0$ depending on the dispersion
relation $\Lambda_{\bm k}$ around its nodes.

According to this classification, which turns out to be
relevant for the study of entanglement, we can
distinguish three phases:
\begin{itemize}
\item  \emph{Phase I}, $\{0\leq\lambda\leq d,\gamma=0\}$,
and $\{\lambda=0,\gamma>0\}$ if $d=2$. For $\gamma=0$, Eq. (\ref{e.Hamilton})
reduces to a simple tight-binding model, which is in a
metallic state with a $2d$-fold symmetric Fermi surface
as far as $\lambda \leq d$.
In $d=2$, for $\lambda=0$ the system is still a metal with a well defined
Fermi surface, which is simply $k_x=k_y \pm \pi$, and whose symmetry
is lowered by the presence of the $\gamma$ term in the Hamiltonian.
In this phase,  $g(0)>0$, and $\bar d = 1$
everywhere except at the point $\{\lambda=d,\gamma=0\}$
where $\bar d = 2$.

\item \emph{Phase II}, $\{0 < \lambda\leq d,\gamma>0\}$,
and  $\{\lambda=0,\gamma>0\}$ if $d=3$.
Away from the boundary lines of this phase,
the system is in a $p$-wave superconducting
state, with a finite pairing amplitude
$\langle c^{\dagger}_{\bm k}c^{\dagger}_{-{\bm k}} \rangle \neq 0$.
Such pairing amplitude vanishes at the boundaries of this
region. The dispersion relation $\Lambda_{\bm k}$ has
point nodes in $d=2$ and line nodes in $d=3$.
Everywhere in this phase $g(0)=0$ and $\bar d = 2$.

\item \emph{Phase III}, $\{\lambda > d\}$. In this phase
the system is in an insulating state with a gap in the
excitation spectrum. Here $g(0)=0$ and $\bar d = d$.
\end{itemize}

This shows that, in terms of the spectral properties,
the above system has \emph{two distinct critical phases},
($I$ and $II$), which are both gapless,
and a \emph{non-critical} phase ($III$).
Numerical evaluation of the
correlation function  through
integration over the first Brillouin zone (FBZ),
\begin{equation}
<c^{+}_{\bm i}c_{\bm j}> = \int_{FBZ}
\frac{d^dk}{(2\pi)^d} \frac{t_{\bm k}}{2\Lambda_{\bm k}}
e^{i{\bm k}\cdot({\bm i}-{\bm j})},
\end{equation}
shows an expected power-law decay in the critical phases
and an exponential decay in the non-critical one.

We then proceed to the evaluation of
the block entropy of entanglement.
The ground state of Eq.(\ref{e.Hamilton}) is known to be a
Gaussian state, whose density matrix can be expressed as the
exponential of a quadratic fermion operator
\cite{Gaudin60,Peschel03}.
To obtain the reduced density matrix of a
$L^d$ subsystem, Grassman algebra is
needed \cite{Peschel01}. Using a Bogoliubov transformation,
the reduced density matrix $\rho_{L}$ can then be written as
\begin{eqnarray}
\rho_{L}=A ~\exp\left(-\sum_{l=1}^{L}\varepsilon_{l}d^{+}_{l}d_{l}\right),
\label{e.rhoL}
\end{eqnarray}
where $d^{+}_{l}$,$d_{l}$ are the new Fermi operators after the
transformation, and $A$ is a normalization constant to ensure
$Tr(\rho)=1$. The single-particle eigenvalues $\varepsilon_{l}$ can
be obtained from $\langle c^{+}_{i}c_{j}\rangle$ and $\langle
c^{+}_{i}c^{+}_{j}\rangle$ by the following
formula\cite{Peschel03}:
\begin{eqnarray}
&(C-F-\frac{I}{2})(C+F-\frac{I}{2})=& \nonumber \\
&\frac{1}{4}~P
~{\rm diag}\left\{\tanh^{2}\left(\frac{\varepsilon_{1}}{2}\right),
\tanh^{2}\left(\frac{\varepsilon_{2}}{2}\right),...,
\tanh^{2}\left(\frac{\varepsilon_{L}}{2}\right)\right\} P^{-1}~~~ &
\end{eqnarray}
where $C_{i,j}=\langle c^{+}_{i}c_{j}\rangle$ and $F_{i,j}=\langle
c^{+}_{i}c^{+}_{j}\rangle$;$P$ is the orthogonal matrix that
diagonalizes the left side of the above equation.
 The Block entropy can then be
calculated in terms of $\varepsilon_{l}$ as:
\begin{eqnarray}
S_L = \sum_{l=1}^{L}\left\{\ln\left[1+\exp(-\varepsilon_{l})\right] +
 \frac{\varepsilon_{l}}{\exp(\varepsilon_{l})+1}
\right \} \label{e.SL}
\end{eqnarray}

\begin{figure}
\includegraphics[width=3.8in]{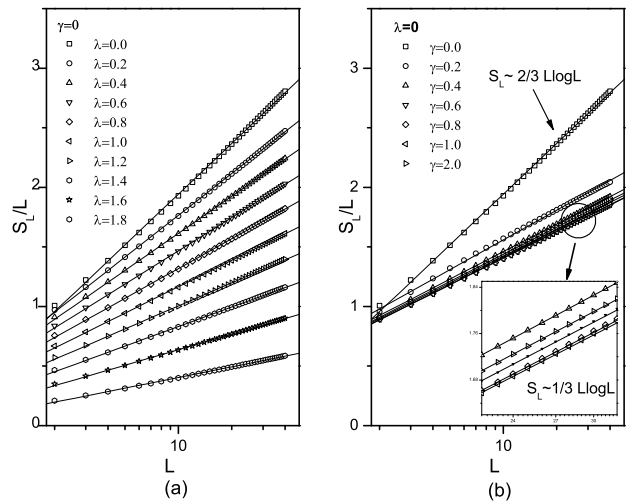}  
\caption{Scaling of the block entropy $S_L$
in $d=2$ for $\gamma=0$ (left panel) and
$\lambda=0$ (right panel). The solid lines correspond to fits
according to the formula  $S_L=\frac{C}{3}L \log_2(L) + B L + A$.}
 \label{f.Gamma0}
\end{figure}

\begin{figure}
\includegraphics[width=3.5in]{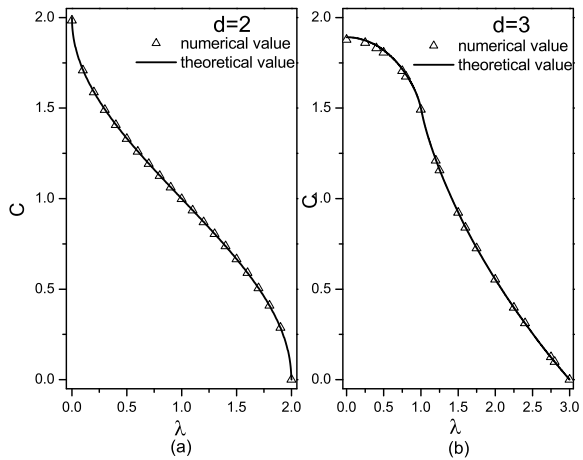}
\caption{$\lambda$-dependence of the $C$ coefficient in
Eq.(\ref{e.logcorr}) in $d=2$ and $d=3$. The values extracted from
fits to our numerical data are compared with the predictions of Ref.
\onlinecite{GioevK05}. In $d=2$, the exact form of $C(\lambda)$ can
be obtained, which is equal to $\frac{2}{\pi}cos^{-1}(\lambda-1)$ }
 \label{f.2D3D}
\end{figure}

 In $d=1$ the above formulas reproduce the scaling of the
 block entropy as observed in the XY model in a transverse
 field \cite{Vidaletal03,Latorreetal04}.
 In $d=2$ the phase diagram is richer, and we need
 to consider the various phases one by one.
 We begin with the critical metallic phase (I), $\gamma=0, 0\leq \lambda < d$.
 For this case a logarithmic correction to the area law,
 $S_L = (C(\lambda)/3) (\log_2 L) L^{d-1}$
 is observed for all values of $\lambda$, as shown in Fig. \ref{f.Gamma0}.
 This is in full agreement with the results of Refs.\cite{Wolf06,GioevK05},
 which predict this behavior in presence of a finite Fermi surface.
 More specifically, Ref. \onlinecite{GioevK05} also supplies us with
 an explicit prediction for the $\lambda$ dependence of $C(\lambda)$,
 based on the Widom's conjecture \cite{Widom81}, in the form
 \begin{equation}
C(\lambda) = \frac1{4 (2\pi)^{d-1}}
\int_{\partial\Omega}\int_{\partial\Gamma(\lambda)} |n_x\cdot n_{p}| dS_{x}
dS_{p} \label{e.widom}
\end{equation}
where $\Omega$ is the volume of the block normalized to one,
$\Gamma(\lambda)$ is the volume enclosed by the Fermi surface,
and the integration is carried over the surface of both domains.
A numerical fit of the calculated asymptotic behavior of $S_L$
through the formula $S_L=\frac{C}{3}L^{d-1} \log_2(L) + B L^{d-1} +
A L^{d-2}+...$
provides us with the
exact result for the $C(\lambda)$ prefactor. In Fig. \ref{f.2D3D}
the prediction of Ref. \onlinecite{GioevK05}, Eq.(\ref{e.widom}),
for the case $\{0\leq \lambda \leq d, \gamma=0\}$ is compared to
our numerical results both for $d=2$ and $d=3$. The agreement is
clearly striking. Moreover, for $\{\lambda=0, \gamma>0\}$
in $d=2$ the formula Eq.(\ref{e.widom}) predicts
$C = 1$, which is also accurately verified by our
data as shown in Fig. \ref{f.Gamma0}.
This proves that the formula Eq.(\ref{e.widom}) is
essentially providing a complete analytic form for the leading behavior
of the block-entropy scaling in arbitrary dimensions for
systems with a finite Fermi surface.

\begin{figure}
\includegraphics[width=3.5in]{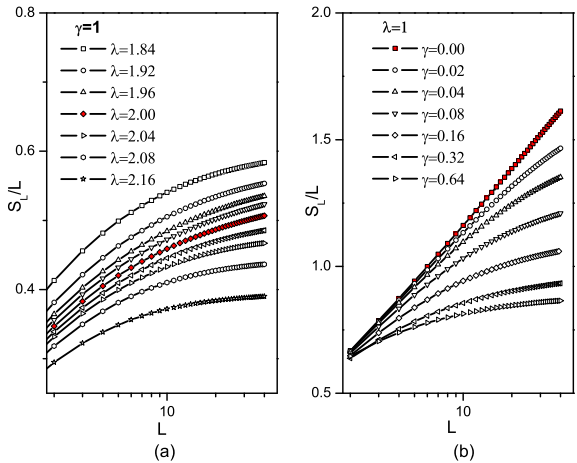}
\caption{Scaling of the block entropy $S_L$ in $d=2$
for $\gamma=1$ (left panel) and $\lambda = 1$ (right panel).}
 \label{f.arealaw}
\end{figure}

 We then turn to the other two phases, $II$ and $III$.
 Two scans through these phases, at fixed $\gamma = 1$ and at
 fixed $\lambda = 1$ are shown in Fig. \ref{f.arealaw}. We observe
 that logarithmic corrections are \emph{absent} in both, and
 only sublogarithmic corrections are possible.
 For $\{\lambda\geq d,\gamma=0\}$ $S_L=0$ identically,
 and the state is not entangled.
  While the area law is expected to hold in the non-critical
  phase $III$, it is surprising to observe it enforced also in
  the critical phase $II$, which has a
  diverging correlation length. This clearly reveals that the
  connection between block-entropy scaling and correlation
  properties is not as straightforward as in $d>1$.

  Our results for the entanglement behavior, co-dimension,
  density of states and correlation properties are
  summarized in Table \ref{t.sum}.
  A crucial difference between the two critical phases
  $I$ and $II$ is the co-dimension $\bar{d}$, and
  the density of states at the ground
  state energy. We have $\bar{d}=1$ and
  $g(0)>0$ in the phase $I$, which shows logarithmic
  corrections to the area law, whereas $\bar{d}=2$
  and $g(0)=0$ in the phase $II$, in which the
  area law is verified up to sublogarithmic corrections.
  It is therefore tempting to conjecture that
  a codimension $\bar{d}\leq 1$ or, alternatively,
  a finite density of states at the ground state
  energy $g(0)>0$ is a \emph{necessary condition} for
  critical phases in $d>1$ to show violations
  of the area law. For the fermionic system under
  consideration, $\bar{d}=1$ requires the existence
  of a finite Fermi surface, which is the basic
  assumption of the proof of area-law violation
  in Refs. \onlinecite{Wolf06,GioevK05}.
  This conjecture would generalize
  the results for $d=1$, where the co-dimension can only
  take the value $\bar{d}=1$, and only critical phases
  with $g(0)>0$ have been explored in the literature.

\begin{table}
\begin{tabular}{|c|c|c|c|c|} \hline
 & $S_L$ & ~$\bar{d}$~ & ~$g(0)$~ & $\langle c_i^{\dagger}c_j \rangle$ \\ \hline
 Phase I & $\sim(\log_2 L) L^{d-1}$ & $1$ & $>0$ & power-law decay \\ \hline
 Phase II & $\sim L^{d-1}$ & 2 & 0 & power-law decay \\ \hline
 Phase III & $\sim L^{d-1}$ & $d$ & 0 & exp. decay \\ \hline
\end{tabular}\caption{Summary of the entanglement scaling properties,
co-dimension, density of states
and decay of correlations in the three phases of
the model Eq.(\ref{e.Hamilton}) in $d=2,3$.}
\label{t.sum}
\end{table}



Further investigations in systems with $d>1$ are clearly needed
to confirm this picture, and to clarify whether more severe
violations of the area law are possible in presence of
infinitely degenerate ground states or
in systems with a fractal Fermi surface \cite{Wolf06}.
During the completion of this manuscript we became aware of
Ref. \onlinecite{Bartheletal06} whose results are in full
agreement with the ones reported in our work.

We thank  A. Cassidy, P. Sengupta and I. Grigorenko for
useful discussions. T.R. acknowledges support of the European
Union through the SCALA project. This work was supported by the
Petroleum Research Foundation, grant ACS PRF$\#$ 41757.



\end{document}